\documentclass[pra,10pt,letterpaper,draft,bibnotes,notitlepage,final,balancelastpage,nofootinbib]{revtex4}

\usepackage{times,bm,bbm,bbold,amssymb,amsbsy,amsthm,amsmath,amsfonts,mathrsfs,graphicx,graphics,color,xcolor,hyperref}
\hypersetup{colorlinks,linkcolor=blue,citecolor=blue,urlcolor=blue}

\begin{document}

\title{Simulation of Quantum Correlation Functions is not Sufficient Resource to Describe Quantum Entanglement}


\author{Akbar Fahmi\footnote{E-mail address: fahmi@theory.ipm.ac.ir}}


\affiliation{Department of Philosophy of Science, Sharif University of Technology, Tehran,14588, Iran}


\date{{\small{\today}}}

\begin{abstract}
The Bell theorem expresses that quantum mechanics is not a local-realistic theory, which is often interpreted as nonlocality of the nature. This result has led to this belief that nonlocality and entanglement are the same resources. However, this belief has been critically challenged in the literature. Here, we reexamine the relation between nonlocality and entanglement in light of the Brassard-Cleve-Tapp (BCT) model, which was originally proposed for simulating quantum correlation of Bell's states by using shared random variables augmented by classical communications. We derive a new criterion for distinguishing quantum mechanics from the BCT model through suggesting an \emph{observable event} based on the perfect correlations (anti-correlations) relation. In particular, we show that in the BCT model one can obtain equal outputs for two opposite input settings with the nonzero probability $0.284$. Hence, in this sense we argue that the BCT model can give rise to an unphysical result. We also show the same problem with a nonlocal version of the BCT model.
\end{abstract}
\maketitle

\section{Introduction}

The quantum entanglement was first described in the Einstein-Podolsky-Rosen theorem (EPR) \cite{EPR}. Einstein, Podolsky and Rosen have imposed locality, reality, and necessary condition for completeness of theory principles (classical concepts) on microscopic world and have found paradoxical result. They pointed out that according to special relativity, violation of locality was impossible and therefore, quantum mechanics must be incomplete. Einstein famously called it ``action at a distance" \cite{act}. Bell has realized the EPR assumptions from the philosophical debates to the experiment. He has proposed mathematics description for EPR assumptions and has derived an inequality which is now known as the Bell inequality \cite{Bell}. He showed that the correlations among the measurement outputs of space-like separated parties on some quantum states cannot be reproduced by realistic local theories. Bell's inequality has been derived in various ways \cite{CHSH,CH}, and many experiments \cite{As,Tit,Han,Sha,Ant,Win} have since been performed that are consistent with quantum mechanical predictions---in conflict with local realism. The violation of Bell inequality usually mean that interpreted as nonlocality of the nature \cite{LMR}.
Entanglement and nonlocality have essential applications in various areas of quantum mechanics, including more recent applications in condensed matter physics \cite{Fa}, quantum thermodynamics \cite{PS}, biology \cite{Wh}, and quantum games \cite{Bu}.


$\vspace{.1cm}$

Quantifying nonlocal correlations created upon measurement on entangled quantum system is an interesting problem in physics and computer sciences. An insightful and natural approach to quantifying nonlocality is to obtain the number of classical bits required to be communicated from one party (conventionally called ``Alice") to the other one (the conventionally called ``Bob"), in addition to using shared random variables in order to simulate quantum correlations. In this direction, simulation of Bell's correlations has recently become an interesting and relevant investigation \cite{Mau,BCT,St,TB,Gisin1,Cerf1}. In this direction, Maudlin \cite{Mau}, Brassard, Cleve, and Tapp \cite{BCT}, and Steiner \cite{St} independently have shown that the simulation can be done with a finite amount of communications. Brassard, Cleve, and Tapp proposed a model that $8$ bits of classical communication suffice for a perfect simulation of the Bell correlation functions. Afterward, Steiner, followed by Gisin and Gisin \cite{Gisin1}, has shown that if one allows the number of bits to vary from one instance to another, then the Bell correlation is reproduced exactly using on average $2$ bits. Besides, Cerf, Gisin and Massar have also shown that if many singlets are to be simulated in parallel, then block coding could be used to decrease the number of communicated bits to $1.19$ bits on average \cite{Cerf1}. Later, Toner and Bacon \cite{TB} have improved these results and have shown that $1$ bit of classical communication is sufficient to reproduce the Bell correlation (for the singlet state). There also exist other studies on simulation of quantum correlations by using instantaneous nonlocal effects, irrespective of the amounts of communicated classical bits \cite{Hall,Non1,De,Mo,GR,BR,Gis,Bell1,Aa,Mer,KS}. In such studies, it has been assumed that Alice's measurements cause instantaneous nonlocal effects on Bob's outputs, where these parties have no access to shared random (hidden) variables.

$\vspace{.1cm}$

The above two structures are identical at the level of what these two frameworks aim to calculate, in other words, e.g., the one-bit classical communication in the Toner---Bacon model \cite{TB} can be replaced with a nonlocal effect. It means that the marginal and joint probabilities calculated in either of these scenarios are the same as those within the other one.

$\vspace{.1cm}$

Nevertheless, there are other viewpoints about relation between quantum entanglement and nonlocality.

$\vspace{.1cm}$

One of them believes that the quantum entanglement and nonlocality are different resources, \cite{BGS}, they showed that the simulation of non-maximally entangled states requires larger amount of nonlocal resources than the simulation of the singlet state. In this direction, Werner \cite{W} introduced a family of highly symmetric, mixed entangled states (``Werner" states), and showed an explicit local hidden-variable model which reproduce, the statistics of quantum measurement when the two parties perform local projective measurements on their qubits. This implies that the Werner states do not violate any Bell's inequality when subjected to projective measurements. Barrett \cite{Ba} extended Werner's model and constructed a new local hidden variable model which is also valid for general measurements performed on some specific entangled Werner states. Recently, it showed that entanglement and nonlocality are inequivalent \cite{TA,Ac,Ac1}, or proved that entanglement, steering, and Bell nonlocality are inequivalent for general measurements \cite{AB}.

$\vspace{.1cm}$

In addition, as mentioned above, the Bell theorem is often interpreted as nonlocality in quantum mechanics. However, it is important to understand which nonlocal models are compatible with quantum mechanics, and which nonlocal models are not. There are some attempts which show inconsistency between nonlocal models and quantum mechanical predictions. For example,  Suarez and Scarani suggested a testable nonlocal hidden variable model \cite{SS} which has later been shown inconsistence with quantum predictions \cite{Zb,Zb1}. Moreover, in different approach and in a series of papers, Leggett and others suggested a specific nonlocal model which is inconsistent with quantum mechanical predictions \cite{Leg}. This model has simulated quantum correlation function \emph{only} for short range of measurement settings. This model has been extended and also realized in experiments \cite{Non,G,Bra,Gi,Ren}. Besides, Jarrett discussed locality assumption and showed that it can be reduce to parameter-independent and outcome-independent conditions \cite{Ja}. There are other efforts to relax outcome-independence condition by allowing an action at the distance from one measurement outcome to the other \cite{Cas}. Their results have indicated the incompatibility of quantum mechanics with a subclass of nonlocal models, which includes Bell local realistic models as a special case.

$\vspace{.1cm}$

As states in above, all aforementioned works have been restricted to very specific mixed states \cite{W} or partially simulated quantum correlation function by nonlocal models \cite{Leg}. In this Paper, we analyze the relation of entanglement and nonlocality from a different perspective by extending the inequivalence between entanglement and nonlocality to \emph{pure maximally Bell states}. We examine one of stepping stone model which simulate pure maximally Bell states \cite{BCT} and show that it is incompatible with perfect correlations (anti-correlations) conservation law. We propose an \emph{observable event} by given one input setting in Alice's site and two opposite input axes in Bob's site. The outputs in the Bob's site correspond to two commutating operators (in the context of quantum mechanics), thus they can be determined and observed simultaneity. These observable quantities in the BCT model are not consistent with predictions of perfect quantum correlations (anti-correlations) law. We find that with probability $0.284$, Bob detects equal outputs for two opposite input axes. This, however, is an unreasonable result. This work can be considered not only as next progress of the Leggett model which is a realistic nonlocal model stronger than Leggett model, but also as next development of Werner \cite{W} states which shows entanglement and nonlocality are inequivalent for pure maximally entangled states. Finally, we derive optimal probability distribution function for various measurement settings of Alice. These results show that true description of nature requires an even more radical modification of our classical notion of locality and reality.





\section{Entanglement and the stochastic BCT model are inequivalent}

\subsection{Bell's correlation}

Quantum correlation for Bell's maximally entangled state $|\phi^{+}\rangle=(1/\sqrt{2})(|00\rangle+|11\rangle)$ has three simple properties: (i) Alice and Bob each measure their qubit's spin along a direction parametrized
by three-dimensional unit vectors $\mathbf{a}$ and $\mathbf{b}$, respectively. Alice and Bob obtain results, $c_{A}\in\{+1,-1\}$ and $c_{B} \in\{+1,-1\}$, respectively, which indicate whether the spin was pointing along ($+1$) or opposite ($-1$)
the direction each party chose to measure. If Alice and Bob's measurement settings be parallel ($\mathbf{a}=\mathbf{b}$) then with probability $1$, the parties get the same outputs ($P(c_{A} = c_{B}|\mathbf{a}=\mathbf{b})=1$). (ii) In each round of the experiment, if Alice (Bob) reverses her (his) measurement axes $\mathbf{a}\rightarrow -\mathbf{a}$ ($\mathbf{b}\rightarrow -\mathbf{b}$), with probability $1$, her (his) outputs are flipped $c_{A}\rightarrow -c_{A}$ ($c_{B}\rightarrow -c_{B}$) as well. And, (iii) the parties joint probability to obtain outputs $c_{A}$ and $c_{B}$, only depends on $\mathbf{a}$ and $\mathbf{b}$ via the combination $P(c_{A}=c_{B}|\mathbf{a}, \mathbf{b})=\cos^{2}\frac{\theta_{a,b}}{2}$, where $\theta_{a,b}$ is the angle between $\mathbf{a} $ and $\mathbf{b}$.

$\vspace{.1cm}$

\emph{Remark 1.} We use Latin alphabet $\mathbf{a},\mathbf{b},\ldots$ for unit vectors showing the direction of measurements. Similarly, we use Greek alphabet  $\alpha,\beta,\gamma,\theta,\ldots$ for angles (as defined later). Besides, to indicate a new unit vector or its associated angle on the unit circle generated from a vector $\mathbf{a}$ by a $\theta$ rotation, we use the shorthand $a+\theta$.

\subsection{Stochastic BCT model}

\begin{figure}
\centering
\includegraphics[scale=.75]{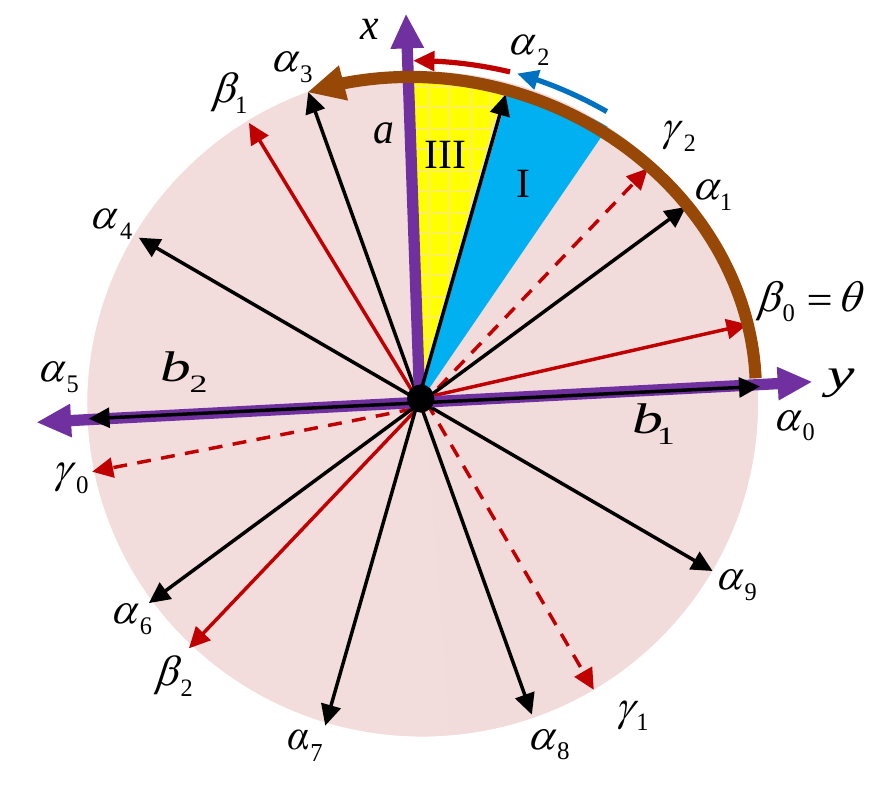}
\caption{(Color online). Random (hidden) variable distribution in the $(\mathbf{x},\mathbf{z})$ plane. Here $\alpha_{j}=j\pi/5$ ($j\in\{0,1,\ldots,9\}$) are ten equally-spaced points on the unit circle. Alice and Bob's measurement settings lie in the $(\mathbf{x},\mathbf{z})$ plane, presented by vectors $\mathbf{a}$ and $\mathbf{b}_{i}$ ($i=1,2$), respectively, such that $\mathbf{a}$ is orthogonal to $\mathbf{b}_{i}$s. Random (hidden) variables are defined as $\beta_{i}$ ($i\in\{0,1,3\}$) for the $(\mathbf{a},\mathbf{b}_{2})$ measurement setting, and $\gamma_{k}$ ($k\in\{0,1,3\}$) for the $(\mathbf{a},\mathbf{b}_{1})$ measurement setting. Here, we select $\alpha_{0}=b_{1}=0$, hence $\theta\in[0, 3\pi/5)$ and $\beta_{0}=\theta$. The brown arc shows variations of $\theta$, from $b_{1}$ to $b_{1}+3\pi/5$.}
\label{BCT1}
\end{figure}

In the BCT model, the maximally entangled state $|\phi^{+}\rangle=(1/\sqrt{2})(|00\rangle+|11\rangle)$ is simulated \cite{BCT}. In this model, the shared random variables are $c \in \{-1,+1\}$,
$\theta \in [0, 3\pi/5)$, and both are uniformly distributed. Here, a parameter $\alpha_{j}$ has also been defined as $\alpha_{j}=j\pi/5$ ($j\in\{0,1,\ldots,9\}$), which indicates ten equally spaced points on the unit circle. The $j$th $\alpha$ slot has been defined as the interval $[\alpha_{j},\alpha_{j+1\hspace{.1cm}\mathrm{mod}10}]$. Moreover, the following quantities also have been defined: $\beta_{0}=\alpha_{0}+\theta, \beta_{1}=\alpha_{3}+\theta,$ $\beta_{2}=\alpha_{6}+\theta$, $\gamma_{0}=\alpha_{5}+\theta, \gamma_{1}=\alpha_{8}+\theta$, and $\gamma_{2}=\alpha_{1}+\theta$ (where the addition is understood to be modulo $2\pi$). Define the $j$th $\beta$ slot as the interval $[\beta_{j},\beta_{j+1\hspace{.1cm}\mathrm{mod}3}]$, and the $j$th $\gamma$ slot as the interval $[\gamma_{j},\gamma_{j+1\hspace{.1cm}\mathrm{mod}3}]$ (Fig. \ref{BCT1}). Alice's and Bob's outputs are the bits $c_{A}$ and $c_{B}$ (here $c_{A},c_{B} \in \{-1,+1\}$), respectively.

$\vspace{.1cm}$

The protocol proceeds as follows:

1- Alice sends the information specifying the $\alpha$-slot, $\beta$-slot, and $\gamma$-slot to Bob, corresponding to where $\mathbf{a}$ is. These slots partition the unit circle into sixteen intervals, hence Alice needs to send $4$ bits of classical information to inform Bob. Alice's outputs are random variables $c_{A}=c$.

After obtaining the information specifying the $\alpha$-slot, $\beta$-slot, and $\gamma$-slot of $\mathbf{a}$ from
Alice, Bob performs the following procedure:

2- If the difference between the $\alpha$-slot numbers of $\mathbf{a}$ and $\mathbf{b}$ is more than two, then Bob sets $b$ to $b + \pi$, $c_{B}$ to $- c_{B}$, and Bob's outputs are $c_{B}=-c_{A}=-c$.

3- If the $\alpha$-slot number of $\mathbf{b}$ is $\in\{7,8,9,0,1\}$, then Bob sets $\beta_0, \beta_1, \beta_2$ to $\gamma_0, \gamma_1, \gamma_2$, respectively.

4- If $\mathbf{a}$ and $\mathbf{b}$ are in the same $\beta$-slot, then Bob takes $c_{B}=c_{A}=c$; otherwise, there exists a $\beta_k$ between $\mathbf{a}$ and $\mathbf{b}$, so Bob defines $u=|b-\beta_k|$, and takes $c_{B}=c_{A}=c$ with probability $1 - (3\pi/10)\sin u$.

5- Alice and Bob repeat the above steps many times.

The BCT model uses the above operations to simulate quantum measurement scenario on Bell's state $|\phi^{+}\rangle$ \cite{BCT}.

\subsection{Entanglement and the stochastic BCT model are inequivalent.}

We consider the original BCT protocol and show that there is an observable event---based on the BCT model---which does not agree with quantum mechanical predictions. To more further clarify our reasoning, we consider one party (Alice) in one site and two parties (we call Bob1 and Bob2) in another site. Bob1 and Bob2 are in the same place and receive the same information from Alice. Alice's input is represented by the three-dimensional unit vector $\mathbf{a}$, and the corresponding outputs are represented by $c_{A}$. Bob1 and Bob2 take their inputs to be in $\mathbf{b}_{1}$ and $\mathbf{b}_{2}$ directions, and the corresponding outputs are represented by $c_{B,1}$ and $c_{B,2}$, respectively. Here, $\mathbf{b}_{1}=-\mathbf{b}_{2}$ and $\mathbf{a}$ is orthogonal to $\mathbf{b}_{1}$ and $\mathbf{b}_{2}$, and they lie in the $(\mathbf{x},\mathbf{z})$ plane (as shown in Fig. \ref{BCT1}). We select $\alpha_{0}=b_{1}=0$, thus $\theta$ changes from $b_{1}$ to $b_{1}+3\pi/5$, and $\beta_{0}=\theta$. Here, Alice, Bob1, and Bob2 run BCT protocol, Bob1 and Bob2 compare their outputs for each rounds of protocol, they obtain, with probability $0.284$, identical outputs for two opposite input settings which physically is unreasonable.

$\vspace{.1cm}$

In the above approach, Bob1 and Bob2 input settings are in two opposite directions.

$\vspace{.1cm}$

\emph{Remark 2.} Here, to more clarify, we considered two Bobs. However, in the BCT protocol context, it doesn't mean that Bob1 and Bob2 perform two sequence of measurements on the second qubit. In fact, we can use one Bob so that he ``calculates" outputs in the $\mathbf{b}$ direction and in the $-\mathbf{b}$ direction.
He compares outputs in the $\mathbf{b}$ and $-\mathbf{b}$ directions for each rounds of protocol. In the context of quantum mechanics, the measurement operators in Bob site correspond to two commuting operators $\bm{\sigma}_{B}\cdot\mathbf{b}$ and $\bm{\sigma}_{B}\cdot(-\mathbf{b})$, and $\bm{\sigma}=(\sigma_{x},\sigma_{y},\sigma_{z})$). Thus, it is not important that Bob performs measurement in the $\mathbf{b}$ or $-\mathbf{b}$ directions.

$\vspace{.1cm}$

\begin{figure}
\centering
\includegraphics[scale=.5]{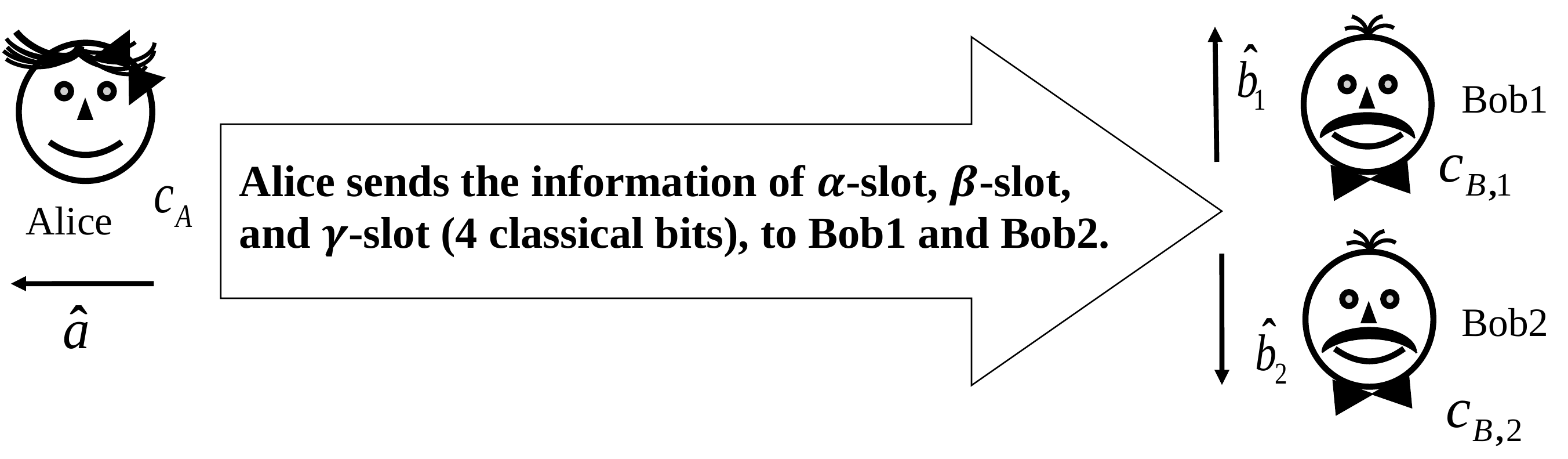}
\caption{Schematic of the BCT protocol. Alice takes her input setting along $\mathbf{a}$ and her output is $c_{A}$. Bob1 and Bob2 are in the same place and receive the same information from Alice. Bob1 and Bob2 take their input settings in $\mathbf{b}_{1}$ and $\mathbf{b}_{2}$ directions ($\mathbf{b}_{1}=-\mathbf{b}_{2}$), and calculate the corresponding outputs which are represented by $c_{B,1}$ and $c_{B,2}$, respectively.}
\label{NBCT}
\end{figure}

As stated by steps (3) and (4) in the subsection B, Bob1 and Bob2 select their random variables and probabilities as follows:

\textbf{\emph{i-1}}- Bob1 uses $\gamma_0, \gamma_1$, and $\gamma_2$ variables for the $\mathbf{b}_{1}$ of the input. Bob1 uses Alice's sent information and denotes the probability of the parties' (Alice and Bob1) outputs for the input settings $(\mathbf{a},\mathbf{b}_{1})$ by $P(c_{A},c_{B,1}|a,b_{1},\theta)$.

\textbf{\emph{i-2}}- Bob2 uses $\beta_0, \beta_1$, and $\beta_2$ variables for the $\mathbf{b}_{2}$ of the input setting. Bob2 uses Alice's sent information and denotes the probability of the parties' (Alice and Bob2) outputs for the input settings $(\mathbf{a},\mathbf{b}_{2})$ by $P(c_{A},c_{B,2}|a,b_{2},\theta)$.

Note that the specific selection of $\alpha_{0}, b_{1}=0$ and $\theta$ do not compromise generality of our conclusion in the sequel.

$\vspace{.1cm}$

\emph{Remark 3.} According to the anti-correlation relation for the two opposite input settings ($\mathbf{b}_{1}=- \mathbf{b}_{2}$) in the Bob1 and Bob2 site, they must obtain opposite outputs for each round of the protocol (for each value of $\theta$).

$\vspace{.1cm}$

Below, we prove that in order to show the existence of an observable event which does not agree with anti-correlation relation for the two opposite input settings, it suffices to consider the random variable $\theta$ in the intervals $\pi/5+\pi/10\leqslant\theta \leqslant 2\pi/5$ and $2\pi/5\leqslant\theta\leqslant \pi/2$.

$\vspace{.3cm}$

\begin{figure}
\centering
\includegraphics[scale=.75]{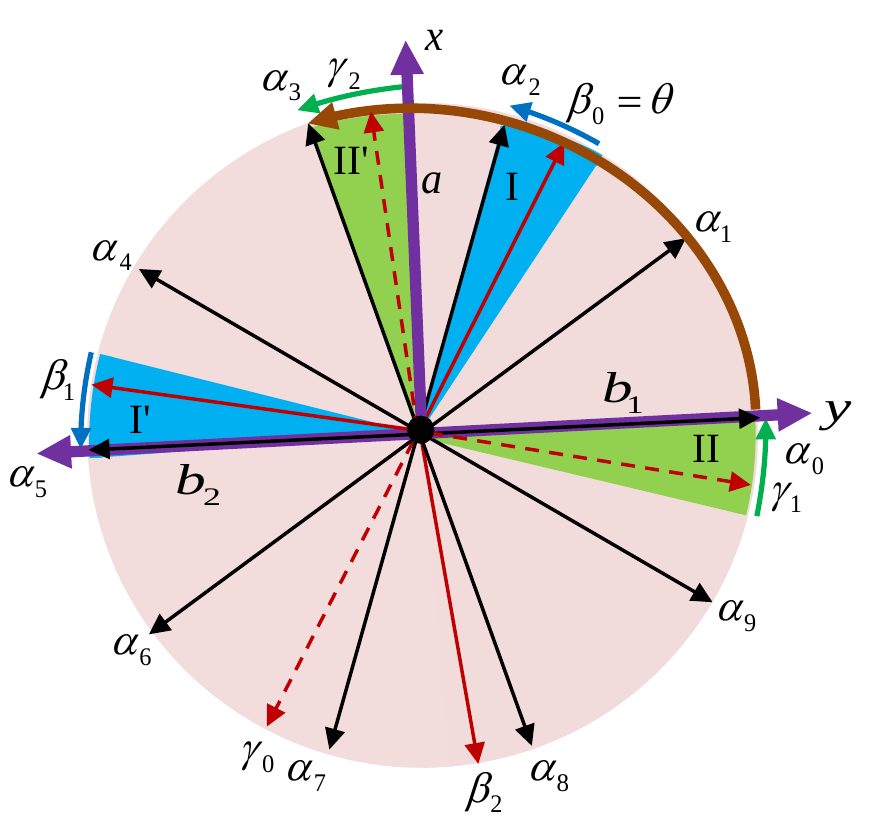}
\caption{(Color online).
Distribution of the hidden variable $\theta$ on the unit circle in the $\pi/5+\pi/10\leqslant\theta \leqslant 2\pi/5$ interval---the blue arc I. For the input setting $(\mathbf{a},\mathbf{b}_{1})$, Alice and Bob obtain equal outputs with probability $P(c_{A}=c_{B,1}|a,b_{1},\theta)=1$ (relative to $\gamma_{i}$s)---the green arcs in the intervals II and II$'$. In the same $\theta$ interval, for the input setting $(\mathbf{a},\mathbf{b}_{2})$, Alice and Bob obtain equal outputs with probability $P(c_{A}=c_{B,2}|a,b_{2},\theta)=1-(3\pi/10)\sin u$ (with $u=|b_{2}-\beta_1|$), relative to $\beta_{i}$s --- the blue arc in the interval I$'$. The probability of obtaining $c_{B,1}=c_{B,2}$  is equal to $P(c_{B,1}=c_{B,2}|b_{1},b_{2})=P(c_{A},c_{B,2}|a,b_{2})\approx0.142$.}
\label{BCT5}
\end{figure}

$\bullet$ In the first interval, $\pi/5+\pi/10\leqslant\theta\leqslant 2\pi/5$ [interval I in Fig. \eqref{BCT5}]:

\textbf{\emph{ii})} Alice sends the information specifying  $\alpha_{2}$-slot, $\beta_{0}$-slot, and $\gamma_{1}$-slot to Bob1 and Bob2 ($4$ bits of classical information).  Alice's outputs are random variables $c_{A}=c$.

\textbf{\emph{iii})} After obtaining the information from Alice, Bob1 and Bob2 perform the following procedure:

\textbf{\emph{iii-1})} Bob1's input setting $\mathbf{b}_{1}$ leis in the $\alpha_0, \beta_2$, and $\gamma_1$ slots. Because $\mathbf{a}$ and $\mathbf{b}_{1}$ are in the same $\gamma$-slot ($\gamma_{1}$-slot), then Bob1 takes $c_{B,1}=c_{A}=c$, with probability equal to one $(P(c_{A}=c_{B,1}|a,b_{1},\theta)=1)$.

\textbf{\emph{iii-2})} Bob2's input setting  $\mathbf{b}_{2}$ lies in the $\alpha_{5}, \beta_1$, and $\gamma_2$ slots. Because $\mathbf{a}$ and $\mathbf{b}_{2}$ are not in the same $\beta$-slot (Alice's slot is $\beta_{0}$), then Bob2 defines $u=|b_{2}-\beta_1|$, and takes $c_{B,2}=c_{A}=c$ with probability $1 - (3\pi/10)\sin u$.

According to the anti-correlation relation for the two opposite measurement settings ($\mathbf{b}_{1}=- \mathbf{b}_{2}$) in the Bob1 and Bob2 site, they must obtain opposite outputs for each round of the protocol (for each value of $\theta$).

Now, Bob1 and Bob2 compare their outputs for each rounds of the protocol. They see their outputs are equal, with probability $P(c_{A}=c_{B,1}=c_{B,2}|a,b_{1},b_{2})=(5/3\pi)\int_{0}^{\pi/10}\left[1-(3\pi/10)\sin u\right]du\approx 0.142$. It means, with probability $0.142$, Bob1 and Bob2 obtain the same outputs for each value of $\theta$ in the first interval.

$\vspace{.1cm}$

\emph{Remark 4.} It is clear from the above calculations that in each run/round of the protocol, Bob1 and Bob2 compare their outputs for a ``specific" shared random variable $\theta$.

$\vspace{.3cm}$

\begin{figure}
\centering
\includegraphics[scale=.75]{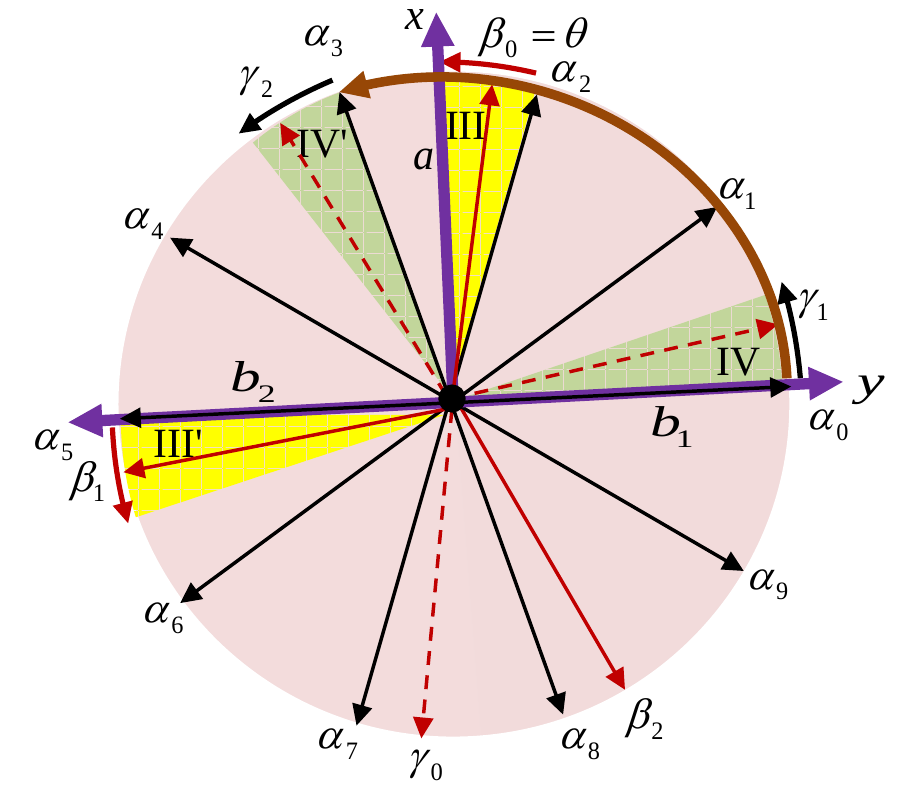}
\caption{(Color online).
Distribution of the hidden variable $\theta$ on the unit circle in the $2\pi/5\leqslant\theta \leqslant \pi/2$ interval---the red arc in the yellow interval III. For the in[ut setting $(\mathbf{a},\mathbf{b}_{2})$, Alice and Bob obtain equal outputs with probability $P(c_{A}=c_{B,2}|a,b_{2},\theta)=1$, relative to $\beta_{i}$s --- the red arcs in the yellow intervals III and III$'$. In the same $\theta$ interval, for the input setting $(\mathbf{a},\mathbf{b}_{1})$, Alice and Bob obtain equal outputs with probability $P(c_{A}=c_{B,1}|a,b_{1},\theta)=1-(3\pi/10)\sin w$ (with $w=|b_{1}-\gamma_1|$), relative to $\gamma_{i}$s---the black arc in the olive interval IV$'$. The probability of obtaining $c_{B,1}=c_{B,2}$ is equal to $P(c_{B,1}=c_{B,2}|b_{1},b_{2})=P(c_{A},c_{B,1}|a,b_{1})\approx0.142$.}
\label{BCT7}
\end{figure}

$\bullet$ In the second interval, $2\pi/5\leqslant\theta \leqslant \pi/2$ [interval III in Fig. \ref{BCT7}]:

\textbf{\emph{iv})} Alice sends the information specifying $\alpha_{2}$-slot, $\beta_{0}$-slot, and $\gamma_{1}$-slot to Bob1 and Bob2. Alice's outputs are random variables $c_{A}=c$.

\textbf{\emph{v})} After obtaining the information from Alice, Bob1 and Bob2 perform the following procedure:

\textbf{\emph{v-1})} Bob1's input setting $\mathbf{b}_{1}$ leis in the $\alpha_0, \beta_2$, and $\gamma_0$ slots. Because $\mathbf{a}$ and $\mathbf{b}_{1}$ are not in the same $\gamma$-slot (Bob1's slot is $\gamma_{0}$, Alice's slot is $\gamma_{1}$), then Bob1 defines $w=|b_{1}-\gamma_{1}|$ and takes $c_{B,1}=c_{A}=c$ with probability $P(c_{A}=c_{B,1}|a,b_{1})=1-(3\pi/10)\sin w$.

\textbf{\emph{v-2})} Bob2's input setting $\mathbf{b}_{2}$ lies in the $\alpha_5, \beta_0$, and $\gamma_2$ slots. Because $\mathbf{a}$ and $\mathbf{b}_{2}$ are in the same $\beta$-slot ($\beta_{0}$-slot), then Bob2 takes $c_{B,2}=c_{A}=c$ with probability equal to one $(P(c_{A}=c_{B,2}|a,b_{2},\theta)=1)$.

Now, Bob1 and Bob2 compare their outputs for each rounds of the protocol. They see their outputs are equal, with probability
$P(c_{A}=c_{B,1}=c_{B,2}|a,b_{1},b_{2})=(5/3\pi)\int_{0}^{\pi/10} \left[1-(3\pi/10)\sin w\right]dw\approx0.142$. It means, with probability $0.142$, Bob1 and Bob2 obtain the same outputs for each value of $\theta$ in the second interval.

As a result, with probability $P(c_{B,1}=c_{B,2}|b_{2}=b_{1}+\pi)=0.284$, Bob1 and Bob2 obtain identical outputs for the two opposite input settings $(\mathbf{b}_{1}=-\mathbf{b}_{2})$, which is physically unreasonable.


$\vspace{.2cm}$

Although, the BCT protocol correctly predicts the correlation function for Alice and Bob1 (with the input settings $(\mathbf{a},\mathbf{b}_{1})$) and for Alice and Bob2 (with the input settings $(\mathbf{a},\mathbf{b}_{2})$) separately, it violates the \textbf{natural principle} that any physical theory should respect.

$\vspace{.3cm}$

\emph{Remark 5.} Although in our approach the input settings $(\mathbf{a},\mathbf{b}_{1})$ and $(\mathbf{a},\mathbf{b}_{2})$ lie in the same $(\mathbf{x},\mathbf{z})$ plane, our methodology is also applicable when the qubits are independently subjected to arbitrary von Neumann measurements not both represented necessarily by vectors in the $(\mathbf{x},\mathbf{z})$ plane (Theorem $3$ of Ref. \cite{BCT}).

$\vspace{.1cm}$

\emph{Remark 6.} In fact, our point on the BCT model is independent from quantum mechanics. We have considered the spin perfect correlation (anti-correlation) conservation law as a fundamental principle which any reasonable physical theory such as classical mechanics, quantum theory, and quantum field theory should respect. This principle used in derivation of realistic models. For example, in the original Bell inequality \cite{Bell}, Bell used this property to derive his inequality. In addition, we would remark that there exists another (apparently trivial) principle, stem from logic, need to be respected by all theories. This principle, called the law of non-contradiction \cite{wiki}, stated that ``contradictory statements cannot be true in the same sense and at the same time". That is, two sets of propositions ($A \& B$) and ($A \& -B$) are mutually exclusive. Having said this, we note that in the BCT model Bob obtains two values of $+1$ and $-1$ simultaneously for the spin of his qubit. Thus the BCT model contradict the law of the non-contradiction too.

$\vspace{.1cm}$

\emph{Remark 7.} It may seem that the above  problem in the BCT model can be fixed if one sets the outcome of $-\mathbf{b}$ to be opposite that of $\mathbf{b}$. Here, we show that the BCT model has internal inconsistency and it can not be fixed by the above simple improvement. This point refer to the spin perfect correlation (anti-correlation) conservation law, which is very close to the second step in the BCT model. The spin perfect correlation (anti-correlation) conservation law can be explained mathematically as follows: The probability to obtain $c_{B}$ for the input setting $b$ is equal to the probability to obtain $- c_{B}$ for the input setting $b+\pi$, for each value of the hidden variable $\theta$ (each round of the BCT protocol);
\begin{eqnarray*}
P(c_{A}=c_{B}|a,b,\theta)=P(c_{A}=-c_{B}|a,b+\pi,\theta), \forall \hspace{.1cm}\theta.
\end{eqnarray*}
However, as we mentioned in the above, this equality is not consistent with other steps of the BCT protocol. For example, in the first interval, $\pi/5+\pi/10\leqslant\theta\leqslant 2\pi/5$, the probability to obtain $c_{A}=c_{B}$ for the input in the $b$ direction is equal to one $P(c_{A}=c_{B}|a,b,\theta)=1$, but the probability to obtain $c_{A}=-c_{B}$ for the another input in the $b+\pi$ direction is equal to  $P(c_{A}=-c_{B}|a,-b,\theta)=(3\pi/10)\sin u$, where $u=|b_{2}-\beta_1|$. In the second interval, $2\pi/5\leqslant\theta \leqslant \pi/2$, the probability to obtain $c_{A}=c_{B}$ for the input in the $b$ direction is equal to $P(c_{A}=c_{B,1}|a,b_{1})=1-(3\pi/10)\sin w$, where $w=|b_{1}-\gamma_{1}|$; whereas the probability to obtain $c_{A}=-c_{B}$ for the input in the $b+\pi$ direction is equal to zero $P(c_{A}=-c_{B}|a,b+\pi,\theta)=0$. It shows that the BCT model has internal inconsistency; it is not consistent with the spin perfect correlation (anti-correlation) conservation law.

\subsection{Entanglement and Nonlocal version of the stochastic BCT model are inequivalent}

Here, we propose a nonlocal version of the BCT protocol which represents an imaginary device which includes two input-output ports, one at Alice's location and the other one at Bob's location, even though Alice and Bob can be space-like separated. This model is one-way protocol so that only Alice's measurement affects Bob's outputs, however, there is no violation of the law of causality implied by the relative time order of events in the space-like regions, and the faster than light communication is imposable. The nonlocal BCT (NBCT) protocol proceeds as follows: the parties share a nonlocal box with hidden variable $\theta \in [0, 3\pi/5)$, $\alpha_{i}$ ($i=0,1,\ldots,9$), $\beta_{j}$, and $\gamma_{k}$ ($j,k=1,2,3$) as in the BCT protocol. Alice selects input along $\mathbf{a}$ and her output is $c_{A}=c$. Alice's operation causes nonlocal effect on Bob's outputs such that if Bob selects his input in the $\mathbf{b}$ direction, his output will be $c_{B}$, as mentioned in the BCT protocol (Fig. \ref{NBCT}). The BCT and NBCT models are similar in the level of what the parties aim to calculate. Thus, in this sense they are equivalent models. In the NBCT model, the classical communications are replaced with nonlocal effects such that the marginal and joint probabilities calculated in either of these protocols is equal to those within the other one. In fact, in the BCT model, the random variables $\theta$, $\alpha$, $\beta$, and $\gamma$ are accessible for Alice and Bob. In the NBCT protocol, we generalized BCT model so that parties have not any access to random variables. However, in what follows we show the existence of an observable event---based on the NBCT (or BCT) model---which does not agree with quantum mechanical predictions.

Similar to previous subsection, it can be shown that there is an observable event---based on the NBCT model---which does not agree with quantum mechanical predictions. Alice and Bob run BCT protocol, Bob obtains, with probability $0.284$, identical outputs for two opposite input settings which physically is unreasonable---to see the details of the calculation, see the Appendix A.

\begin{figure}
\centering
\includegraphics[scale=.75]{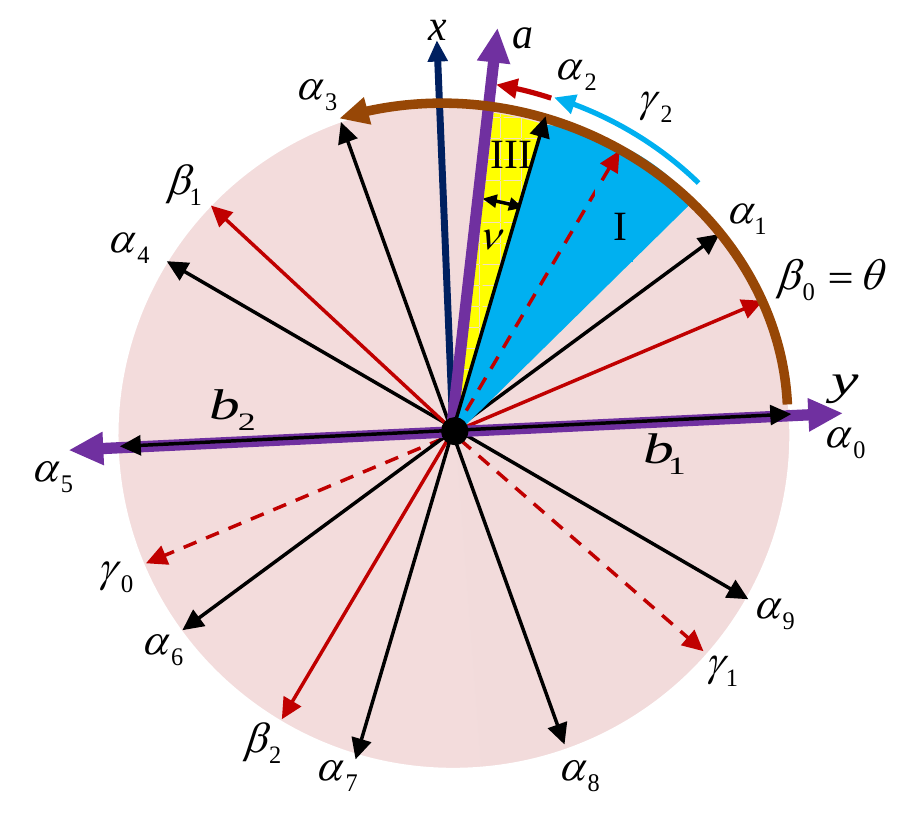}
\caption{(Color online). Distribution of (the hidden variable) $\theta$ on the unit circle for obtaining optimal probability of unacceptable outputs. The direction of the unit vector $\mathbf{a}$ can change from $\bm{\alpha}_{2}$ to $\bm{\alpha}_{3}$. The angle between vectors $\bm{\alpha}_{2}$ and $\mathbf{a}$ is defined as $\nu$ ($0\leqslant\nu\leqslant\pi/5$). For the subset of the hidden variable within the $\pi/5+\nu\leqslant\theta \leqslant 2\pi/5$ interval---the blue arc, in the blue interval I---and the $2\pi/5\leqslant\theta\leqslant 2\pi/5+\nu$ interval---the red arc in the yellow interval III---with probability $P(c_{B,1}=c_{B,2}|b_{1},b_{2},\nu)=\left[-\frac{2}{3}+\frac{1}{2}(\cos\nu+\cos(\frac{\pi}{5}-\nu))\right]$, Bob obtains identical outputs for two opposite input settings ($\mathbf{b}_{1}=-\mathbf{b}_{2}$), which is physically unreasonable.}
\label{BCT11}
\end{figure}

In what follows, we briefly investigate the optimal probability to obtain inconsistent outputs and the probability distributions in a real experimental scenario in the NBCT model.

\section{Optimal probability to obtain inconsistent outputs}

Let us restrict, without loss of generality, that Alice's input setting is associated to the $a\in[\alpha_{2}, \alpha_{3}]$ interval. Besides, we denote the angle between the vectors $\bm{\alpha}_{2}$ and $\mathbf{a}$ by $\nu$, where (evidently) $0\leqslant\nu\leqslant\pi/5$ (as shown in Fig. \ref{BCT11}). It can be shown that for Bob to obtain an inconsistent output (contradicting quantum mechanics predictions)---namely to obtain identical results for inputs on two opposite axes---the probability becomes $P(c_{B,1}=c_{B,2}|b_{1},b_{2},\nu)=\left[-\frac{2}{3}+\frac{1}{2}(\cos\nu+\cos(\frac{\pi}{5}-\nu))\right]$. This probability obtains its maximum value at $\nu=\pi/10$, with $P_{\max}(c_{B,1}=c_{B,2}|b_{1},b_{2},\nu=\frac{\pi}{10})=0.284$, and the minimum occurs at $\nu=0, \pi/5$, with $P_{\min}(c_{B,1}=c_{B,2}|b_{1},b_{2}, \nu=0, \frac{\pi}{5})=0.071$. The details of calculations and figures can be found in the Appendix B.

\section{Probability distributions in a real experimental scenario}

The probability distribution functions in actual experimental setups are typically reduced by a visibility factor $V$, due to, e.g., imperfections in the source, measurement apparatus, and decoherence. Hence, the total probability to obtain identical outputs (for the two opposite measurement settings) is modified as  $V^{2}\left[-\frac{2}{3}+\frac{1}{2}(\cos\nu+\cos(\frac{\pi}{5}-\nu))\right]$, which is positive for all values of $V$. Therefore, one should be able to distinctively detect our results in experiments. However, if we assume existence of an intelligent device such that it does not allow identical outputs to appear, one still can find a visibility threshold $V_{\mathrm{th}}=53.99\%$ (at $\nu=\pi/10$) so that for $V_{\mathrm{exp}}>V_{\mathrm{th}}$ there exists a non-vanishing probability to obtain identical outputs for two opposite measurement settings in Bob's site. The details of the calculations can be found in the Appendix C.

\section{Summary}

It is generally accepted that entanglement and nonlocality are equivalent. However, there is another viewpoint which advocates that there are different resources. These works have been restricted to specific mixed states or non-maximally entangled states \cite{TA,Ac,Ac1}. Moreover, Leggett has shown an incompatibility between quantum mechanics and a class of nonlocal realistic models. This work stimulated a number of theoretical and experimental works from different perspectives \cite{Non,G,Bra,Gi,Ren}. Besides, there are other efforts to relax local causality assumptions of the Bell inequality \cite{Cas}. They demonstrated, both in theory and experiment, a conflict between their model and quantum predictions and observed measurement data.


$\vspace{.1cm}$

In this article, for the first time, we have analyzed the relation of entanglement and nonlocality by extending the inequivalence between them within the scope of \emph{pure maximally-entangled states}. We have considered a nonlocal model which exactly simulates quantum correlation functions, and we have shown that it is not compatible with the spin perfect correlation (anti-correlation) conservation law. Here, we list our main results:

$\vspace{.1cm}$

A- We have rigorously shown that one of the stepping-stone articles (arguing the relation between quantum entanglement and nonlocal models) has internal inconsistency. This defect is not artificial, it is fundamental.

$\vspace{.1cm}$

B- It is generally accepted that entanglement and nonlocality are equivalent, however, there is another viewpoint which believe they have different resources. These works have been restricted to very specific mixed states or non-maximally entangled states. Here, for the first time, we have analyzed the relation of entanglement and nonlocality from a different perspective by extending the inequivalence between them to \emph{pure maximally Bell states}.


$\vspace{.1cm}$

C- The BCT model is considered as rudimentary for other some of the papers \cite{Cs,Bra1}. Thus any serious comment on the rigor of the BCT model will have formulated import on the message and validity of other derivative results.

$\vspace{.1cm}$

D- In all previous works, it was believed that quantum entanglement has been simulated by some communicating models. However, we have shown that quantum entanglement has \emph{other} intrinsic properties which any model shall respect. For example, spin vector satisfy the perfect correlation (anti-correlation) relation; spin vectors transform under unitary operators, such as rotation to another spin vectors. Now, this important question arises that how such properties are translated in the language of hidden variable models.

$\vspace{.1cm}$

E- It shows that simulating quantum correlation function by using shared random (hidden) variable models which augmented by classical communications (nonlocal effects) is a \emph{necessary} feature but it is not \emph{sufficient}. There are other physical properties which need to be tested.

$\vspace{.1cm}$

$\vspace{.2cm}$

\textbf{Acknowledgments:} We thank A. T. Rezakhani for discussions.


\section{Appendix}

In what follows, we show entanglement and nonlocality are inequivalent in nonlocal version of the stochastic BCT model in detail. Moreover, we derive the probabilities in the perfect and unperfect cases; and an explicit description of the optimal probability value for the measurement along the two opposite axes in Bob's site. Finally, we find a Visibility threshold in a real experiment scenario, and show there exists a non-vanishing probability to violate one of the properties of Bell's correlations.

\subsection{Entanglement and nonlocality are inequivalent in Nonlocal version of the stochastic BCT model.}

\begin{figure}
\centering
\includegraphics[scale=.5]{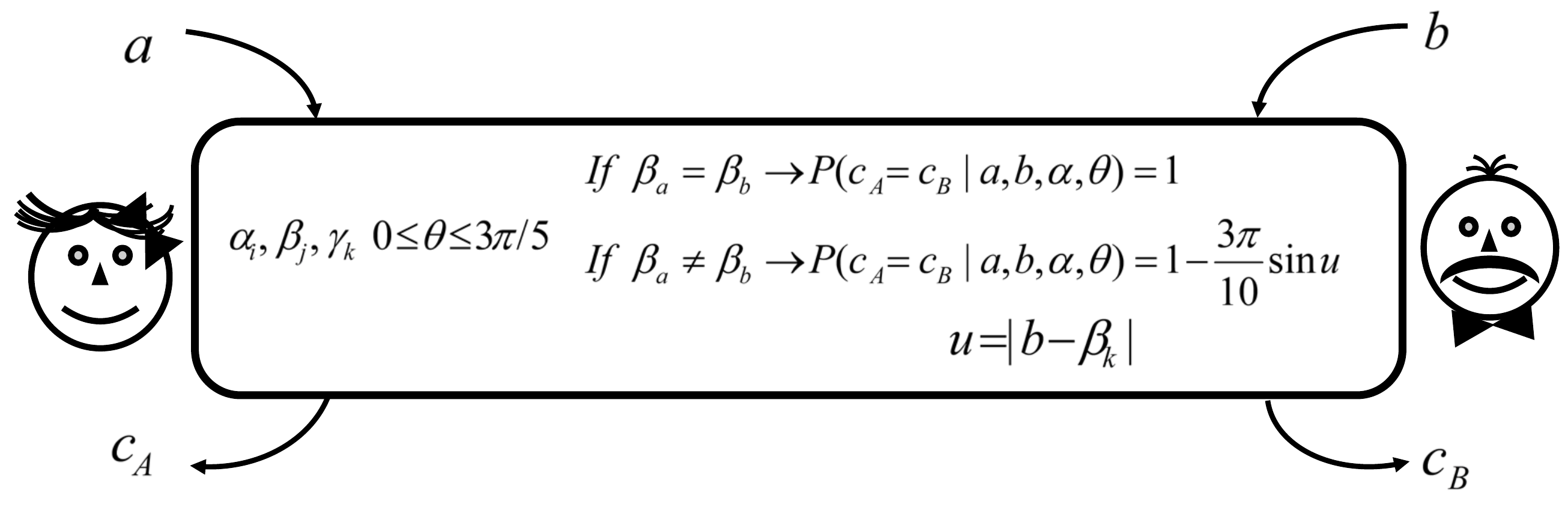}
\caption{Schematic of the NBCT protocol. Here, $\alpha_{j}$ ($j\in\{0,1,\ldots,9\}$), $\beta_{i}$ ($i\in\{0,1,3\}$), $\gamma_{k}$ ($k\in\{0,1,3\}$), and $\theta \in [0, 3\pi/5)$ are hidden variables in the NBCT protocol. The Alice input is in the $\mathbf{a}$ direction and her output is $c_{A}$. Alice's operation causes a nonlocal effect on Bob's output such that if Bob selects $\mathbf{b}$ direction as input, his output will be $c_{B}$, as mentioned in the BCT protocol.}
\label{NBCT}
\end{figure}

We take two sets of input settings $(\mathbf{a},\mathbf{b}_{1})$ and $(\mathbf{a},\mathbf{b}_{2})$ in the $(\mathbf{x},\mathbf{y})$ plane (as shown in Fig. \ref{BCT1}), and the corresponding outputs $(c_{A},c_{B,1})$, and $(c_{A},c_{B,2})$, respectively. We select $\alpha_{0}=b_{1}=0$, thus $\theta$ changes from $b_{1}$ to $b_{1}+3\pi/5$, and $\beta_{0}=\theta$. As stated by steps (3) and (4) of the BCT model, $\gamma_0, \gamma_1$, and $\gamma_2$ variables are used for the $\mathbf{b}_{1}$ of the $(\mathbf{a},\mathbf{b}_{1})$ input setting; similarly, $\beta_0, \beta_1$, and $\beta_2$ variables are used for the $\mathbf{b}_{2}$ of the $(\mathbf{a},\mathbf{b}_{2})$ input setting. Note that the specific selection of $\alpha_{0}, b_{1}=0$, and $\theta$ do not compromise generality of our conclusion in the sequel. We denote the probability of the parties' outputs for the input settings $(\mathbf{a},\mathbf{b}_{1})$ and $(\mathbf{a},\mathbf{b}_{2})$ by $P(c_{A},c_{B,1}|a,b_{1})$ and $P(c_{A},c_{B,2}|a,b_{2})$, respectively.


We consider input settings as shown in Fig. \ref{BCT5}. As stated by steps (3) and (4) of the BCT model, $\gamma_0, \gamma_1$, and $\gamma_2$ variables are used for $\mathbf{b}_{1}$ of the $(\mathbf{a},\mathbf{b}_{1})$ input setting; similarly, $\beta_0, \beta_1$, and $\beta_2$ variables are used for $\mathbf{b}_{2}$ of the $(\mathbf{a},\mathbf{b}_{2})$ input setting.

$\vspace{.1cm}$

To obtain observable event which is not consistent with anti-correlation outputs in the BCT model, it is sufficient to consider two subsets of the hidden variable $\theta$ in the intervals $\pi/5+\pi/10\leqslant\theta \leqslant2\pi/5$ and $2\pi/5\leqslant\theta\leqslant \pi/2$.

$\vspace{.1cm}$

$\bullet$ In the first interval, $\pi/5+\pi/10\leqslant\theta\leqslant 2\pi/5$ [interval I in Fig. \ref{BCT5}], and for the input setting $(\mathbf{a},\mathbf{b}_{1})$, with probability equal to one (i.e., $P(c_{A}=c_{B,1}|a,b_{1},\theta)=1$), Alice and Bob obtain the same outputs (relative to $\gamma_{i}$s). In the same interval, for the input setting $(\mathbf{a},\mathbf{b}_{2})$, the parties obtain the same outputs with probability $P(c_{A}=c_{B,2}|a,b_{2},\theta)=1-(3\pi/10)\sin u$, where $u=|b_{2}-\beta_{1}|$ (relative to $\beta_{i}$s). Thus, for the two inputs in the Bob's site, we have:
\begin{eqnarray}\label{S}
P(c_{A}=c_{B,1}=c_{B,2}|a,b_{1},b_{2})=(5/3\pi)\int_{0}^{\pi/10}\left[1-(3\pi/10)\sin u\right]du\approx 0.142,
\end{eqnarray}
taken over interval I$'$ in Fig. \ref{BCT5}. In other words,

$\frac{\pi}{5}+\frac{\pi}{10}\leqslant\theta \leqslant \frac{2\pi}{5}
\Longrightarrow\left\{
  \begin{array}{ll}
    P(c_{A}=c_{B,1}|a,b_{1},\theta,\gamma_{1})=1, & \hbox{for} \hspace{.3cm} (\mathbf{a},\mathbf{b}_{1}) \\
    P(c_{A}=c_{B,2}|a,b_{2},\theta,\beta_{1})=1-(3\pi/10)\sin u, & \hbox{for} \hspace{.3cm} (\mathbf{a},\mathbf{b}_{2})
  \end{array}
\right.,\\\hbox{from which} \hspace{.1cm} P(c_{A}=c_{B,1}=c_{B,2}|a,b_{1},b_{2},\theta) =P(c_{A}=c_{B,2}|a,b_{2},\theta)$.
This shows that with probability equal to $P(c_{B,1}=c_{B,2}|b_{1},b_{2})=P(c_{A},c_{B,1}|a,b_{1})\approx 0.142$, the quantities $c_{B,1}$ and $c_{B,2}$ are equal.

$\vspace{.1cm}$

$\bullet$ Similarly, in the second interval, $2\pi/5\leqslant\theta \leqslant \pi/2$ [interval III in Fig. \ref{BCT7}], and for the input setting $(\mathbf{a},\mathbf{b}_{2})$, with probability equal to one (i.e., $P(c_{A}=c_{B,2}|a,b_{2},\theta)=1$), Alice and Bob obtain the same outputs (relative to $\beta_{i}$s). In the same interval, and for the input setting $(\mathbf{a},\mathbf{b}_{1})$, the parties obtain the same outputs with probability $P(c_{A}=c_{B,1}|a,b_{1})=1-(3\pi/10)\sin w$, where $w=|b_{1}-\gamma_{1}|$ (relative to $\gamma_{i}$s). Thus, for the two inputs in the Bob's site, we have:
\begin{eqnarray}\label{1}
P(c_{A}=c_{B,1}=c_{B,2}|a,b_{1},b_{2})=(5/3\pi)\int_{0}^{\pi/10} \left[1-(3\pi/10)\sin w\right]dw\approx0.142,
\end{eqnarray}
taken over interval IV in Fig. \ref{BCT7}. In other words,

$\frac{2\pi}{5}\leqslant\theta \leqslant \frac{\pi}{2}
\Longrightarrow\left\{
  \begin{array}{ll}
 P(c_{A}=c_{B,1}|a,b_{1},\theta,\beta_{1})=1-(3\pi/10)\sin w, & \hbox{for} \hspace{.3cm} (\mathbf{a},\mathbf{b}_{1})\\
 P(c_{A}=c_{B,2}|a,b_{2},\theta,\gamma_{1})=1, & \hbox{for;} \hspace{.3cm} (\mathbf{a},\mathbf{b}_{2})
  \end{array}
 \right.,\\\hbox{from which} \hspace{.1cm} P(c_{A}=c_{B,1}=c_{B,2}|a,b_{1},b_{2},\theta) =P(c_{A}=c_{B,1}|a,b_{1},\theta)$. This shows that with probability equal to $P(c_{B,1}=c_{B,2}|b_{1},b_{2})=P(c_{A},c_{B,1}|a,b_{1})\approx0.142$, the quantities $c_{B,1}$ and $c_{B,2}$ are equal.


Hence, with nonzero probability $P(c_{B,1}=c_{B,2}|b_{2}=b_{1}+\pi)=0.284$, Bob obtains identical outputs for the two opposite input settings $(\mathbf{b}_{1}=-\mathbf{b}_{2})$. This is, however, unacceptable noting condition (ii) in Bell's correlation.

\subsection{An upper bound on the optimal probability value for measurement along two opposite axes in the Bob's site}

Here, we derive an upper bound on the optimal probability value for obtaining identical results for input along two opposite axes in Bob's site. Considering step (2) of the BCT protocol, if $\mathbf{a}$ and $\mathbf{b}_{2}$ differ $\geqslant3\pi/5$, Bob's outputs (for the setting with the two opposite input vectors, $\mathbf{b}_{1}=-\mathbf{b}_{2}$) will be consistent with quantum mechanics prediction. Thus, here we restrict ourselves to the case $\alpha_{2}\leqslant a \leqslant\alpha_{3}$, for which inconsistency with quantum mechanical predictions is seen. Let $\nu$ be the angle between vectors $\bm{\alpha}_{2}$ and $\mathbf{a}$, which satisfies $0\leqslant\nu\leqslant\pi/5$ (as shown in Fig. 3, at the main text).

$\vspace{.1cm}$

In order to obtain optimal probability value of an observable event which does not agree with quantum mechanical predictions, it suffices to consider $\theta$ in the intervals $b_{1}+\pi/5+\nu\leqslant\theta \leqslant b_{1}+2\pi/5$ and $b_{1}+2\pi/5\leqslant\theta \leqslant b_{1}+2\pi/5+\nu$.

$\vspace{.1cm}$

\begin{figure}
\centering
\includegraphics[scale=.75]{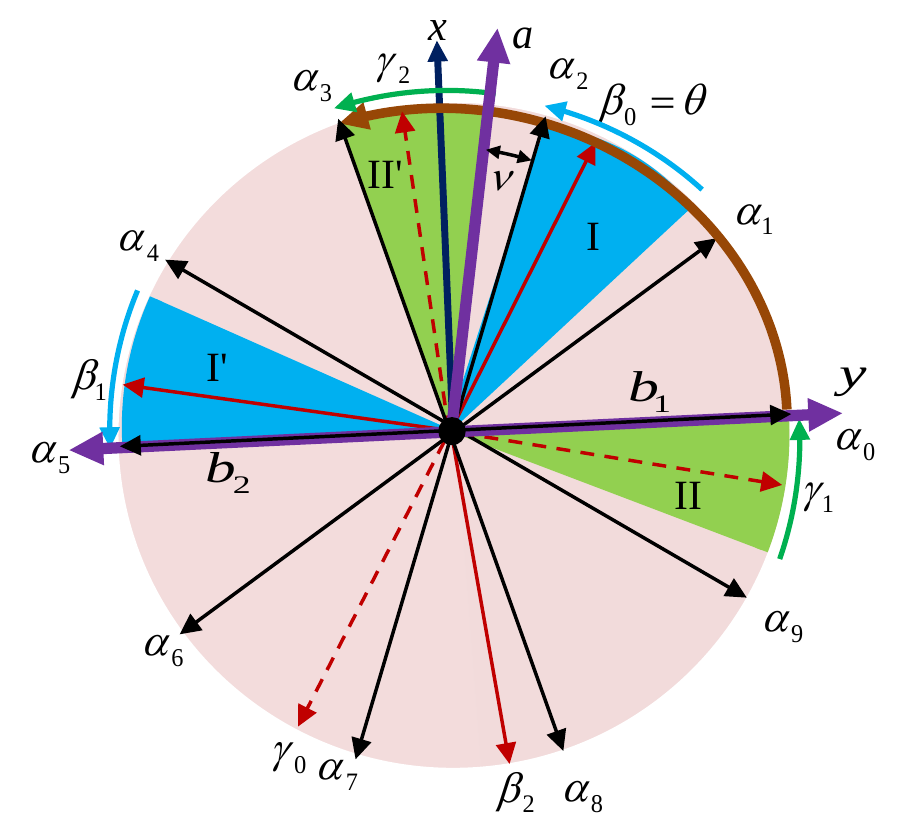}
\caption{(Color online). For the subset of the hidden variable within the $\pi/5+\nu\leqslant\theta \leqslant 2\pi/5$ interval---the blue arc I, in the blue interval I---and for the input setting $(\mathbf{a},\mathbf{b}_{1})$, with probability equal to one, Alice and Bob obtain equal outputs (with respect to $\gamma_{i}$)---the green arcs II and II$'$. In the same hidden variable interval, for the input setting $(\mathbf{a},\mathbf{b}_{2})$, with probability equal to $(5/3\pi)\left[(\pi/5)-\nu-(3\pi/10)(1-\cos(\frac{\pi}{5}-\nu))\right]$, Alice and Bob obtain equal outputs (with respect to $\beta_{i}$s)--- the blue arcs I and I$'$.}
\label{BCT12}
\end{figure}

$\bullet$ In the interval $b_{1}+\pi/5+\nu\leqslant\theta\leqslant b_{1}+2\pi/5$ (interval I in Fig. \ref{BCT12}), and for the input setting $(\mathbf{a},\mathbf{b}_{1})$, Alice and Bob obtain the same outputs (relative to $\gamma_{i}$s), with probability $1$; and for the $(\mathbf{a},\mathbf{b}_{2})$ input setting, the parties obtain the same outputs with probability $P(c_{A}=c_{B,2}|a,b_{2},\theta)=1-(3\pi/10)\sin u$, where $u=|b_{2}-\beta_{1}|$ (relative to $\beta_{i}$s). After taking integral over the aforementioned interval, we obtain
\begin{eqnarray*}
P_{1}(c_{B,1}=c_{B,2}|b_{1},b_{2},\nu)=\frac{5}{3\pi}\int_{0}^{\frac{\pi}{5}-\nu}\left[1-\frac{3\pi}{10}\sin u\right]du=\frac{5}{3\pi}\left[\frac{\pi}{5}-\nu-\frac{3\pi}{10}\big(1-\cos(\frac{\pi}{5}-\nu)\big)\right],
\end{eqnarray*}
where $u=|b_{2}-\beta_{1}|$ (interval I$'$ in Fig. \ref{BCT12}).
\begin{figure}
\centering
\includegraphics[scale=.75]{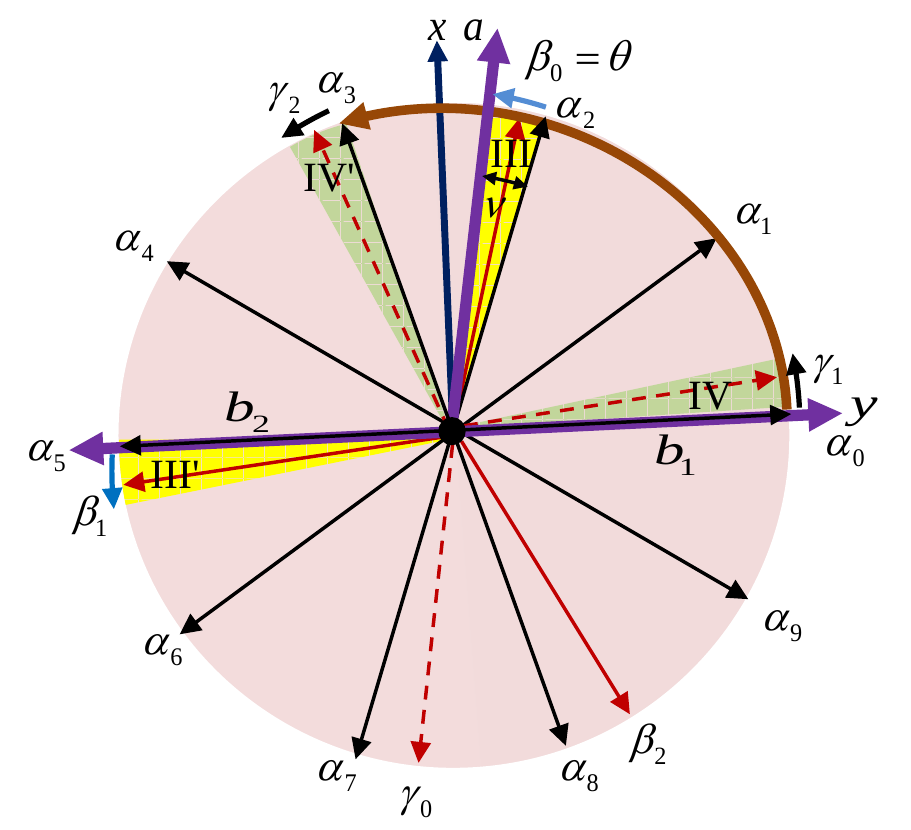}
\caption{(Color online). For the subset of the hidden variable $\theta$ in the $2\pi/5\leqslant\theta\leqslant 2\pi/5+\nu$ interval --- the blue arc in the yellow interval III---and for the input setting $(\mathbf{a},\mathbf{b}_{2})$, with probability equal to one, Alice and Bob obtain equal outputs (respect to $\beta_{i}$s) --- the blue arc in intervals III and III$'$. In the same hidden variable interval, and for the input setting $(\mathbf{a},\mathbf{b}_{1})$, with probability $(5/3\pi) \left[\nu-(3\pi/10)(1-\cos\nu)\right]$ (with respect to $\gamma_{i}$s)---the black arc in the olive interval IV and IV$'$.}
\label{BCT13}
\end{figure}

$\vspace{.1cm}$

$\bullet$ In the interval $2\pi/5\leqslant\theta\leqslant 2\pi/5+\nu$ (interval III in Fig. \ref{BCT13}), and for the input setting $(\mathbf{a},\mathbf{b}_{2})$, Alice and Bob obtain the same outputs (relative to $\beta_{i}$s), with probability $1$; and for the $(\mathbf{a},\mathbf{b}_{1})$ input setting, they obtain the same outputs with probability $P(c_{A}=c_{B,1}|a,b_{1})=1-(3\pi/10)\sin w$ (relative to $\gamma_{i}$s), where $w=|b_{1}-\gamma_{1}|$. After taking integral over the aforementioned interval, we obtain
\begin{eqnarray*}
P_{2}(c_{B,1}=c_{B,2}|b_{1},b_{2},\nu)=\frac{5}{3\pi}\int_{0}^{\nu} \left[1-\frac{3\pi}{10}\sin w\right]dw=\frac{5}{3\pi} \left[\nu-\frac{3\pi}{10}(1-\cos\nu)\right],
\end{eqnarray*}
where $w=|b_{1}-\gamma_{1}|$ (interval IV in Fig. \ref{BCT13}).

\begin{figure}
\centering
\includegraphics[scale=.6]{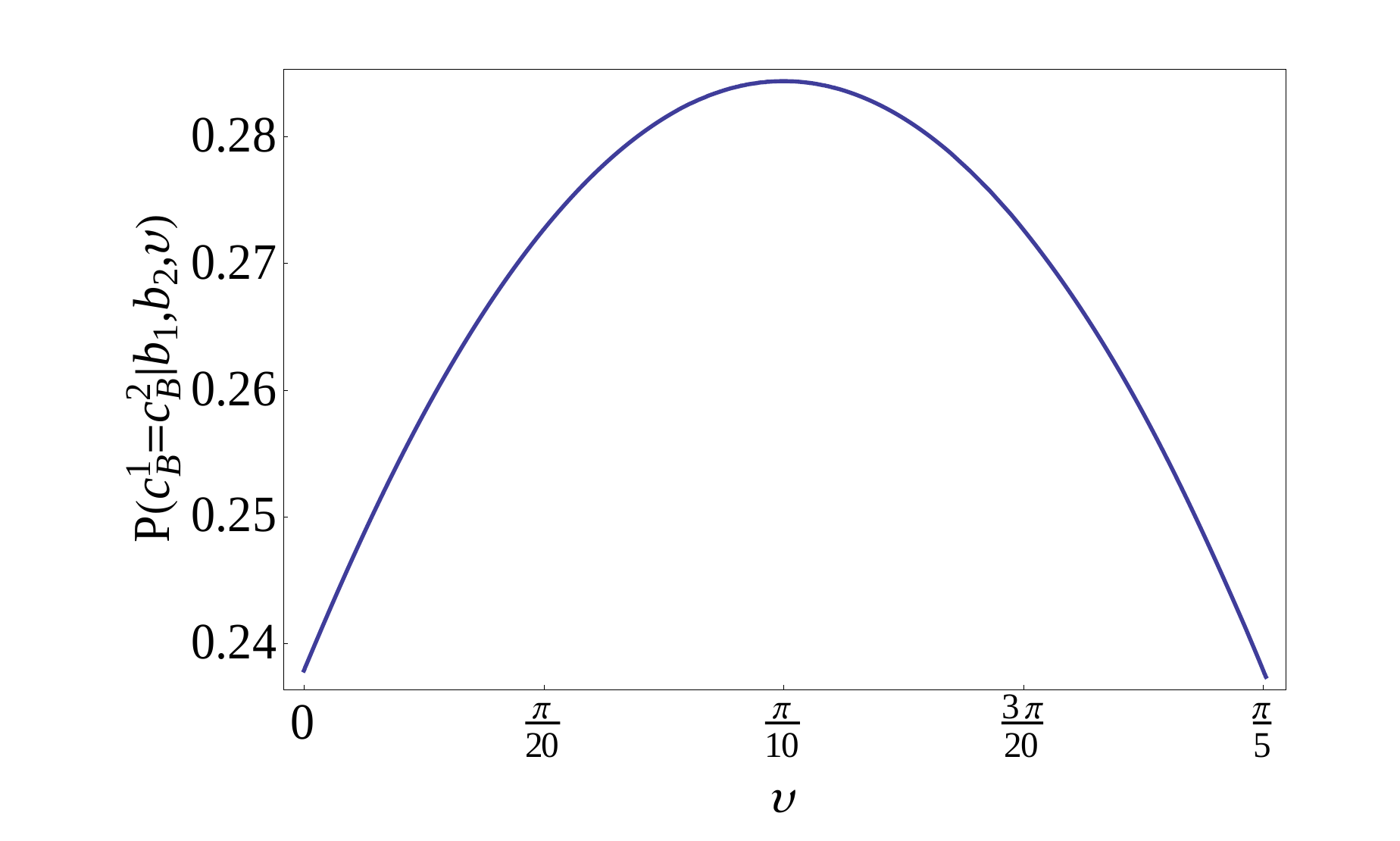}
\caption{(Color online). The probability distribution $P(c_{B,1}=c_{B,2}|b_{1},b_{2},\nu)=\left[-\frac{2}{3}+\frac{1}{2}(\cos\nu+\cos(\frac{\pi}{5}-\nu))\right]$, depicted as a function of $\nu$. The maximum value occurs at the $\nu_{\max}=\pi/10$ where $P_{\max}(c_{B,1}=c_{B,2}|b_{1},b_{2})=0.284$; and the minimum value is at $\nu_{\min}=0, \pi/5$ where $P_{\min}(c_{B,1}=c_{B,2}|b_{1},b_{2})=0.071$.}
\label{BCT14}
\end{figure}

Hence, with probability $P(c_{B,1}=c_{B,2}|b_{1},b_{2},\nu)=P_{1}+P_{2}=\left[-\frac{2}{3}+\frac{1}{2}(\cos\nu+\cos(\frac{\pi}{5}-\nu))\right]$, Bob obtains identical outputs for the two opposite input settings. This is, however, physically unreasonable because of condition (ii) in Bell's correlation. The maximum value of this probability takes place at $\nu=\pi/10$, where $P_{\max}(c_{B,1}=c_{B,2}|b_{1},b_{2},\nu=\frac{\pi}{10})=0.284$ (introduced in the earlier section), and the minimum takes place at $\nu=0, \pi/5$, where $P_{\min}(c_{B,1}=c_{B,2}|b_{1},b_{2}, \nu=0, \frac{\pi}{5})=0.071$ (Fig. \ref{BCT14}).

\subsection{Probability distributions and finding a Visibility threshold in a real experiment scenario.}

Here, we briefly discuss how to observe our results in a realistic experimental scenario. The probability distribution functions in actual experimental setups are typically reduced, from the perfect (`per') values $P_{\mathrm{per}}$ to efficient (`eff') values, by a visibility factor $V$ as $P_{\mathrm{eff}}=VP_{\mathrm{per}}$. This reduction occurs usually because of imperfections in the source, measurement apparatus, and decoherence. Hence, the total probability to obtain identical outputs (for the two opposite input settings) are given by $P_{\mathrm{eff}}(c_{B,1}=c_{B,2}|b_{1},b_{2},\nu)=V^{2}\left[P_{1}+P_{2}\right]=V^{2}\left[-\frac{2}{3}+\frac{1}{2}(\cos\nu+\cos(\frac{\pi}{5}-\nu))\right]$, which is positive for all values of $V$. For example, at the theoretical optimal point $\nu=\pi/10$, given the visibility $99.0\pm1.2\hspace{.2cm}\%$ for measurement in the $H/V$ (horizontal/vertical polarization) basis, the visibility $99.2\pm1.6\hspace{.2cm}\%$ for measurement in the $\pm\pi/4$ basis, and the visibility $98.9\pm1.7\hspace{.2cm}\%$ for measurement in the $R/L$ basis \cite{Non}, we obtain the values $0.278$, $0.279$, and $0.277$, for the corresponding probability. Therefore, one should be able to distinctively detect our results in experiments.

$\vspace{.1cm}$

We remark that in the above discussion, we have implicity assumed independence of the two probability distributions [that is why we multiplied them to obtain $P_{\mathrm{eff}}(c_{B,1}=c_{B,2}|b_{1},b_{2},\nu)$]. If, however, we assume existence of an intelligent device such that it does not allow identical outputs to appear, we find a visibility threshold $V_{\mathrm{th}}=53.99\hspace{.2cm}\%$ (at $\nu=\pi/10$) and show for a visibility factor greater than it, $V_{\mathrm{exp}}>V_{\mathrm{th}}$, there exists a non-vanishing probability to violate one of the properties of Bell's correlations. The visibility factor reduces the probability values from the perfect (`per') value $P_{\mathrm{per}}$ to efficient (`eff') outputs $P_{\mathrm{eff}}=VP_{\mathrm{per}}$ so that the value of inefficient (`inef') outputs becomes $P_{\mathrm{inef}}=P_{\mathrm{per}}-VP_{\mathrm{per}}$, which is not reliable and hence should be discarded.
\begin{figure}
\centering
\includegraphics[scale=.8]{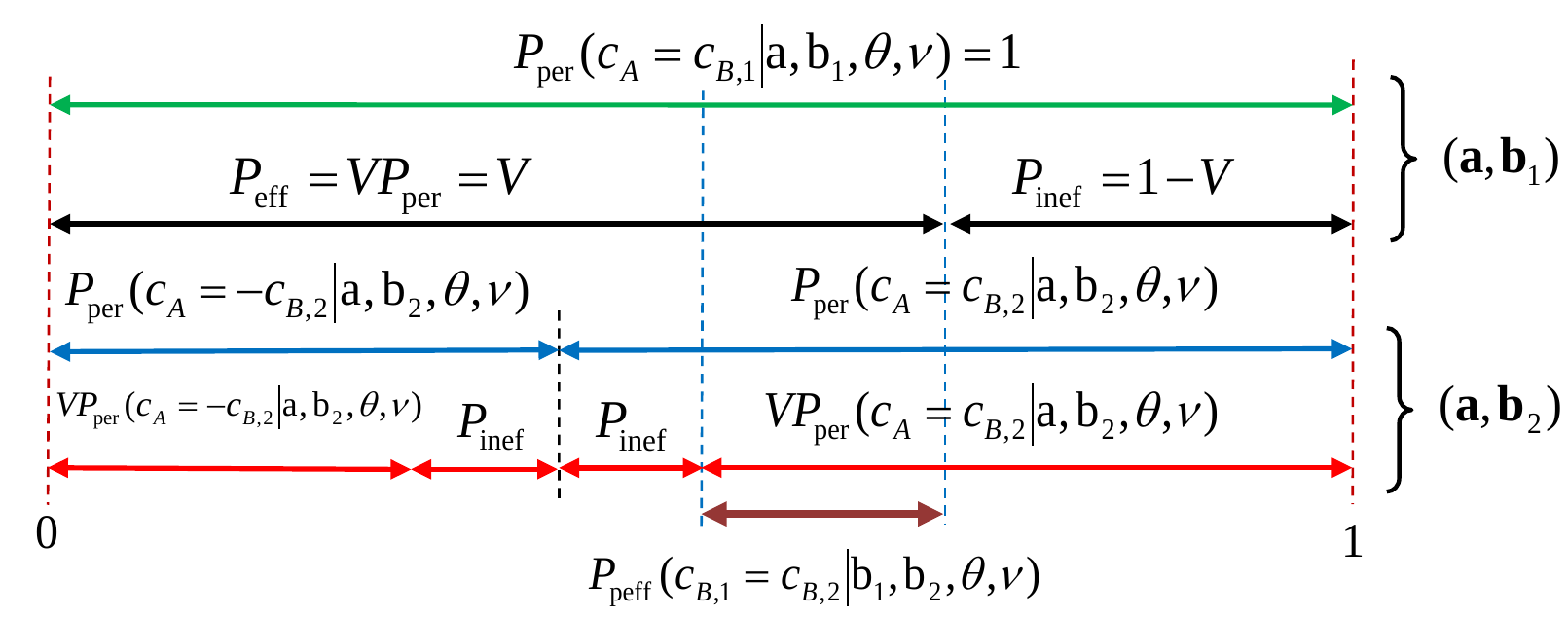}
\caption{(Color online) A schematic of probability values under visibility effects for a specific value of $\theta$ in the interval $b_{1}+\pi/5+\nu\leqslant\theta\leqslant b_{1}+2\pi/5$.}
\label{unf}
\end{figure}
Here, for example, in the interval $\pi/5+\nu\leqslant\theta\leqslant 2\pi/5$ (interval I in Fig. \ref{BCT12}), the perfect probability $P_{\mathrm{per}}(c_{A}=c_{B,1}|a,b_{1},\nu)=1$ (the green line in Fig. \ref{unf}) reduces to $P_{\mathrm{eff}}(c_{A}=c_{B,1}|a,b_{1},\nu)=V$ (the black line), and the value of inefficient outputs becomes $P_{\mathrm{inef}}=1-VP_{\mathrm{per}}$. For the $(\mathbf{a},\mathbf{b}_{2})$ input setting, the perfect probability $P_{\mathrm{per}}(c_{A}=c_{B,2}|a,b_{2},\nu)$ (the blue line in Fig. \ref{unf}), reduces to $P_{\mathrm{eff}}(c_{A}=c_{B,2}|a,b_{2},\nu)=V\frac{5}{3\pi}\left[\frac{\pi}{5}-\nu-\frac{3\pi}{10}(1-\cos(\frac{\pi}{5}-\nu))\right]$ (the red line), and the value of inefficient outputs becomes $P_{\mathrm{inef}}=P_{\mathrm{per}}(c_{A}=c_{B,2}|a,b_{2},\nu)-VP_{\mathrm{per}}(c_{A}=c_{B,2}|a,b_{2},\nu)$. Therefore, if the values of $P_{\mathrm{eff}}(c_{A}=c_{B,1}|a,b_{1},\nu)$ and $P_{\mathrm{eff}}(c_{A}=c_{B,2}|a,b_{2},\nu)$ (the brown line in Fig. \ref{unf}) have common interval (partial efficient outputs (`peff')), we have nonzero probability for inputs along the two opposite axes in Bob's site. Thus, the partial efficient probability is given by
\begin{eqnarray}
P_{\mathrm{peff},1}(c_{B,1}=c_{B,2}|a,b_{1},b_{2},\nu)=V\frac{5}{3\pi}\left[\frac{\pi}{5}-\nu-\frac{3\pi}{10}(1-\cos(\frac{\pi}{5}-\nu))\right]-\frac{(\pi/5)-\nu}{3\pi/5}(1-V). \end{eqnarray}
Similarly for the interval $2\pi/5\leqslant\theta\leqslant 2\pi/5+\nu$ (interval III in Fig. \ref{BCT13}), the partial efficient probability is obtained as
\begin{eqnarray}
P_{\mathrm{peff},2}(c_{B,1}=c_{B,2}|a,b_{1},b_{2},\nu)=V\frac{5}{3\pi} \left[\nu-\frac{3\pi}{10}(1-\cos\nu)\right]-\frac{\nu}{3\pi/5}(1-V).
\end{eqnarray}
Thus, the total partial efficient probability is
\begin{eqnarray*}
&&P_{\mathrm{peff},1}(c_{B,1}=c_{B,2}|a,b_{1},b_{1},\nu)+P_{\mathrm{peff},2}(c_{B,1}=c_{B,2}|a,b_{1},b_{2},\nu)\\
&&\hspace{6.5cm}=V\left[-\frac{2}{3}+\frac{1}{2}(\cos\nu+\cos(\frac{\pi}{5}-\nu))\right]-2\frac{\pi/10}{3\pi/5}(1-V).
\end{eqnarray*}
This means that the efficient probability to obtain identical outputs (for the two opposite measurement settings) is  \begin{eqnarray*}
P_{\mathrm{peff}}(c_{B,1}=c_{B,2}|b_{1},b_{2},\nu)=\max\left\{0,V\left[-\frac{2}{3}+\frac{1}{2}(\cos\nu+\cos(\frac{\pi}{5}-\nu))\right]-2\frac{\pi/10}{3\pi/5}(1-V)\right\}.
\end{eqnarray*}
Thus evidently, for $P_{\mathrm{peff}}(c_{B,1}=c_{B,2}|b_{1},b_{2},\nu)\geqslant0$, we obtain a visibility threshold $V_{\mathrm{th}}$. For example, the visibility threshold is equal to $V_{\mathrm{th}}=53.99\hspace{.2cm}\%$ at $\nu=\pi/10$ in which $P_{\mathrm{exp}}(c_{B,1}=c_{B,2}|b_{1},b_{2},\nu)=0$. If visibility factor $V_{\mathrm{exp}}$ be greater than a visibility threshold $V_{\mathrm{th}}$, there exists a non-vanishing probability to violate one of the properties of Bell's correlations.

\end{document}